\documentclass[prd,12pt,nofootinbib]{revtex4-2}
\usepackage[margin=1in]{geometry}
\usepackage[compat=1.1.0]{tikz-feynman}
\usepackage{subcaption}
\usepackage{amsmath}
\usepackage{amsthm}
\usepackage{mathtools}
\usepackage{commath}
\usepackage{graphicx}
\usepackage{amssymb}
\usepackage{bm}
\usepackage{slashed}
\usepackage{tikz}
\usepackage{braket}
\usepackage[labelsep=space]{caption}

\def\be{\begin{equation}}
\def\ee{\end{equation}}	
\def\bea{\begin{eqnarray}}
\def\eea{\end{eqnarray}}

\def\beq{\begin{equation}}
\def\eeq{\end{equation}}

\def\rar{\rightarrow}

\def\ba{\begin{array}}
	\def\ea{\end{array}}

\usepackage{hyperref}
\usepackage{epstopdf}
\usepackage{comment}

\begin{document}
\title{Constraining the parameter space of a singlet down-type vector-like quark}

\author{Mohammad Ahmady}

\email{mahmady@mta.ca}
\affiliation{\small Department of Physics, Mount Allison University, \mbox{Sackville, New Brunswick, Canada, E4L 1E6}}
\author{Maryam Bibi}
\email{mbibi@mta.ca}
\affiliation{\small Department of Physics, St. Francis Xavier University, \mbox{Antigonish, Nova Scotia, Canada, B2G 2X5}}
\affiliation{\small Department of Physics, Mount Allison University, \mbox{Sackville, New Brunswick, Canada, E4L 1E6}}
%\author{Zoe McIntyre}
%\email{zxmcintyre@mta.ca}
%\affiliation{\small Department of Physics, Mount Allison University, \mbox{Sackville, New Brunswick, Canada, E4L 1E6}}
%\author{Michael Thibodeau}
%\email{methibodeau@mta.ca}
%\affiliation{\small Department of Physics, Mount Allison University, \mbox{Sackville, New Brunswick, Canada, E4L 1E6}}
%\author{Ryan MacGibbon}
%\email{ryan_macgibbon@me.com}
%\affiliation{\small Department of Physics, Acadia University,
%	\mbox{Wolfville, Nova-Scotia, Canada, B4P 2R6}}
\author{Ruben Sandapen}
\email{ruben.sandapen@acadiau.ca}
\affiliation{\small Department of Physics, Acadia University,
	 \mbox{Wolfville, Nova-Scotia, Canada, B4P 2R6}}
\affiliation{\small Department of Physics, Mount Allison University, \mbox{Sackville, New Brunswick, Canada, E4L 1E6}}

\begin{abstract}
We consider a minimal New Physics scenario in which the Standard Model is extended by singlet down-type vector-like quark(VLQ). We constrain its parameter space using the measured branching ratios for $B_s \to \mu^+ \mu^-$, $B \to X_s  \mu^+ \mu^-$ and $B \to X_s \gamma$, which are all consistent with Standard Model predictions. Besides considering the tree-level flavor-changing-neutral-current amplitude allowed in this NP scenario, we account for a Higgs--VLQ loop amplitude and show that their resulting interference drives the VLQ mass into the discovery reach via direct searches at the LHC. We also examine the implications of this minimal NP scenario for the rare $B \to K \nu \bar{\nu}$ decay where there is a persistent tension between the  SM prediction and the Belle II measurements.

\end{abstract}


\maketitle
%%%%%%%%%%%%%%%%%%%%%%%%%%%%%%%%%%%%%%%%%%

\section{Introduction}

The persistent anomalies \cite{Hurth:2023Review,Fischer:2022HiddenPhysics} in rare $B$ decays motivate the exploration of New Physics scenarios \cite{Buras:2014fpa}, one of them being the addition of vector-like quarks to the SM \cite{Alves:2023ufm,Okada:2012VLQ,Vatsyayan:2020VLQmixing}. In the simplest scenario, the SM is augmented by a singlet down-type vector-like quark \cite{Ahmady:2001VLQ,Aoki:2000VLQgamma,Chang:1999VLQ,Morozumi:2004VLQ,Alok:2016VLQ}. Such a vector-like quark allows the $b \to s$ FCNC transition at tree-level, quantified by a coupling $U_{sb}=|U_{sb}|e^{i \theta_{sb}}$ with $|U_{sb}| \propto 1/M_D^2$ where $M_D$ is the VLQ mass. 

Recent direct searches at the LHC, at a center-of-mass energy of $13~{\rm TeV}$, have set lower limits on the masses of VLQs:  $M_U> 1.26~{\rm TeV}$ and $M_D> 1.33~{\rm TeV}$ for the up-type and down-type vector quarks respectively \cite{ATLAS:2022tla}. Assuming that VLQs interact mainly with the third generation of SM quarks, they can be produced at LHC either in pairs or a single VLQ plus a third generation quark.  The latter becomes dominant for larger VLQ mass \cite{Banerjee:2024zvg}.  Production of VLQs are followed by their decay into SM particles.  For example a produced down-type VLQ can decay to $Zb$, $Hb$ or $Wt$.  One can also consider more exotic VLQs decays involving (other) beyond SM particles \cite{Benbrik:2019zdp}. The LHC discovery reach for the production of  singlet VLQs is limited to $M_D \lesssim 3~\text{TeV}$ \cite{AguilarSaavedra:2010rx,Sirunyan:2019sza,ATL-PHYS-PUB-2018-032,CMS-PAS-FTR-18-005}. Conservatively, there remains a window, $1.5~\mathrm{TeV} \lesssim M_D \lesssim 3~\mathrm{TeV}$, for the potential discovery of a VLQ at the LHC.

A legitimate question is whether a singlet down-type VLQ is already ruled out since the rare $B$ decays, $B_s \to \mu^+ \mu^-$, $B \to X_s  \mu^+ \mu^-$ and $B \to X_s \gamma$, are all in agreement with SM predictions. This question has most recently been addressed in Refs. \cite{Alok:2016VLQ} and \cite{Morozumi:2018cnc}, both considering only the tree-level FCNC amplitude (which is proportional to $1/M_D^2$) in addition to the SM amplitude. For this reason, these studies only provide a lower bound on $M_D$. Thus, the VLQ minimal scenario cannot be ruled out: a large enough $M_D$ always remains a possible explanation for the current agreement between SM predictions and experiment. 

In this paper, we go beyond Refs. \cite{Alok:2016VLQ} and \cite{Morozumi:2018cnc} by accounting also for the  Higgs--VLQ loop amplitude. In contrast with the tree-level FCNC amplitude, the Higgs--VLQ loop amplitude grows logarithmically with $M_D$. This means that it starts to dominate the tree-level amplitude as $M_D$ increases. As we shall see, the interference between the SM and total (tree $+$ loop) VLQ amplitude leads to an upper bound on $M_D$. Interestingly, for a safely allowed violation of the CKM unitarity,  this upper bound lies within the direct discovery reach of the LHC.

	\section{An additional down-type VLQ}
    \label{SM+VLQ}
The NP scenario we consider here is the addition of a single down-type VLQ. This VLQ, unlike the SM quarks, has both its left-handed and right-handed components as singlets of $SU(2)_L$.  As a result, the addition of bilinear mass terms like
\be
\mathcal{L}^{\mathrm{Dirac}}_{\mathrm{VLQ}}=M(\bar{D}_RD_L+\bar{D}_LD_R)\;,
\label{diracmass}
\ee
where $M$ is the bare mass of the VLQ, does not break the electroweak gauge symmetry, thus keeping the gauge structure of the SM intact \cite{Ahmady:2001VLQ}. After electroweak symmetry breaking by the Higgs mechanism, the VLQ mixes with SM quarks (with charge $-1/3$) via Yukawa couplings proportional to the Higgs vacuum expectation value, leading to modified quark mixing matrices and potentially observable deviations from SM predictions in flavor-changing processes and precision electroweak observables.

On the other hand, SM quarks acquire their mass through invariant Yukawa couplings ($f^{ij}_{u/d}$) to an isodoublet scalar Higgs field $\phi$:
\be
-f^{ij}_d\bar{\psi}^i_Ld^j_R\phi-f^{ij}_u\bar{\psi}^i_Lu^j_R\tilde{\phi}+{\rm H.C.}\;,
\label{yukawacouplings}
\ee
where the three generations of the SM quarks are indexed as $i,j=1,2,3$ with the quark and Higgs doublets are defined as:
\be
\psi^i_L=
{\begin{pmatrix}
	u^i \\
	d^i \\
\end{pmatrix}}_L,\;\; 
\phi=
{\begin{pmatrix}
		\chi^+ \\
		\frac{1}{\sqrt{2}}(v+H+i\chi^\circ ) \\
\end{pmatrix}},\;\;
\tilde{\phi}=
{\begin{pmatrix}
		\frac{1}{\sqrt{2}}(v+H-i\chi^\circ ) \\
		-\chi^- \\
\end{pmatrix}}\;.
\ee
$H$ and $\chi^{\circ\pm}$ are physical and unphysical Higgs and $v$ is the Higgs vacuum expectation value.  Additionally, an $SU(2)_L$ invariant Yukawa coupling between the down-type VLQ and SM quarks written as
\be
-f^{i4}_d\bar{\psi^i_L}D_R\phi +{\rm H.C.}\;,
\label{DYukawaterm}
\ee
allows mixing among all four down-type quarks.  The non-zero $v$ leads to spontaneous electroweak symmetry breaking and the creation of mass terms for quarks as follows
\be
\bar{d}^\alpha_L\mathcal{M}^{\alpha\beta}_dd^\beta_R 	+\bar{u}^i_L\mathcal{M}^{ij}_uu^j_R+{\rm H.C.}\;,
\ee	
where $\alpha ,\;\beta =1,2,3,4$, and $d_{L,R}^4=D_{L,R}$.  Note that $M_d$ is now a $4\times 4$ matrix which can be written as
\be
\mathcal{M}_d=\frac{v}{\sqrt{2}}
{\begin{pmatrix}
		f_d^{11}&f_d^{12}&f_d^{13}&f_d^{14} \\
		f_d^{21}&f_d^{22}&f_d^{23}&f_d^{24} \\
		f_d^{31}&f_d^{32}&f_d^{33}&f_d^{34}  \\
		0&0&0& \frac{\sqrt{2}M}{v}\\
\end{pmatrix}}\;,
\label{massmatrix}
\ee
where $M$ is the Lagrangian mass of the VLQ.  To diagonalize the mass matrix, $4\times 4$ unitary transformations $A^d_L$ and $A^d_R$ rotate the weak left- and right-handed eigenstates into the mass (physical) eigenstates, i.e.
\be
{A^d_L}^\dagger \mathcal{M}_dA^d_R={\rm diag}(m_d,m_s,m_b,M_D)\;.
\ee
 Since $M\gg v$, $\mathcal{M}_d^{44}$ is the dominant entry of the mass matrix, which means that the mixing between SM quarks and the VLQ is small. Therefore, one can treat the diagonalization perturbatively such that the left-handed mixing can be written as:
\be
{(A^d_L)}^{i4}\approx f^{i4}\frac{v}{\sqrt{2}M}\;.
\label{lefthandedmixing}
\ee
On the other hand, the extended $3\times 4$ CKM quark mixing matrix is the product of the elements of the unitary matrices  $A^u_L$ ($3\times 3$) and $A^d_L$ ($4\times 4$ ):
\be
V^{i\alpha}=\sum_{j=1}^3{A^u_L}^{ij}{A^d_L}^{j\alpha}\;.
\label{CKM}
\ee
One can immediately see that the extended quark mixing matrix Eq. \ref{CKM} is not unitary as \cite{Ahmady:2001VLQ}
\be
{(V^\dagger V)}^{\alpha \beta}=\delta^{\alpha\beta}-{({A^d_L}^*)}^{4\alpha}{A^d_L}^{4\beta}\;.
\ee
This non-unitarity leads to the possibility of having non-zero FCNC at the tree level in the VLQ NP scenario.  For example, the tree level vertices $bsZ$, $bsH$ and $bs\chi^\circ$ are all proportional to the model parameter $U^{sb}$ defined as
\be
U^{sb}=(V^\dagger V)^{sb}=-{({A^d_L}^*)}^{42}{A^d_L}^{43}\approx \frac{f^{42}f^{43}v^2}{M_D^2}\equiv \frac{y_{sb}v^2}{M_D^2}\;,
\ee
where $y_{sb}=f^{42}f^{43}=|y_{sb}|e^{i\theta_{sb}}$. Thus, we have $3$ free parameters in this VLQ scenario: $|y_{sb}|, \theta_{sb}$ and $M_D$. 

\section{Deviation from CKM unitarity}

 Rare B decays, driven by the FCNC $b\rar s$ transition, are heavily suppressed in the SM. This is because FCNCs are forbidden at tree-level and are suppressed by the  Glashow–Iliopoulos–Maiani (GIM) mechanism \cite{PhysRevD.2.1285} at one-loop. Both are consequences of the unitarity of the Cabibbo-Kobayashi-Maskawa (CKM) quark mixing matrix \cite{Cabibbo:1963yz,Kobayashi:1973fv} in the SM. However, experimental tests of the CKM unitary currently allow for a small violation.  
 
 In this section, we calculate the upper bound of the CKM unitarity violation, that is the maximum value of 
 \begin{equation}
\left| \frac{U_{sb}}{V_{tb}V_{ts}^*}\right|=\left|1+ \frac{V_{cb} V_{cs}^*}{V_{tb} V_{ts}^*} +\frac{V_{ub} V_{us}^*}{V_{tb} V_{ts}^*}\right| \;,
\label{nonunitarity}
\end{equation}
 where we use the available experimental data for the CKM matrix elements on the right hand side.  The latter are obtained without imposing the unitarity requirement \cite{ParticleDataGroup:2024}.
 The magnitude of the first ratio on the right-hand side of Eq. (\ref{nonunitarity}) is 
\begin{equation}
\left| \frac{V_{cb} V_{cs}^*}{V_{tb} V_{ts}^*} \right| =0.956 \pm 0.043\;,
\label{vcbvcs}
\end{equation}
as determined from direct measurements of the magnitudes of the individual CKM element:
\[
|V_{cb}| = 0.0411 \pm 0.0012, \quad
|V_{cs}| = 0.975 \pm 0.006, \quad
|V_{tb}| = 1.010 \pm 0.027
\]
(the last one from single top-quark production), and
\[
|V_{ts}| = 0.0415 \pm 0.0009
\]
from $B_s^0$--$\bar{B}_s^0$ mixing mass difference $\Delta m_s$ using lattice QCD inputs and the measured $|V_{tb}|$ \cite{ParticleDataGroup:2024}. The phase of the ratio is determined from the world-average CP-violating phase 
\[
\phi_s = -0.040 \pm 0.016~\text{rad}
\]
in $B_s^0$--$\bar{B}_s^0$ mixing (measured via time-dependent CP asymmetry in $B_s^0 \to J/\psi\,\phi$ and related golden modes such as $B_s^0 \to J/\psi\,\pi^+\pi^-$):
\[
\arg\left( \frac{V_{cb} V_{cs}^*}{V_{tb} V_{ts}^*} \right) = \frac{\phi_s}{2}.
\]

For the second ratio on the right-hand-side of Eq.(\ref{nonunitarity}), using the PDG magnitudes for the matrix elements without imposing unitarity \cite{ParticleDataGroup:2024} and the tree-level determination of the angle $\gamma$, we find
\begin{equation}
 \frac{V_{ub} V_{us}^*}{V_{tb} V_{ts}^*}  =(0.0204 \pm 0.0013)e^{i(115\pm 3)^\circ}\;,
\label{vubvus}
\end{equation}
 leading to the following experimental limit on unitarity violation:

 \begin{equation}
\left| \frac{U^{sb}}{V_{tb}V_{ts}^*}\right| =0.04\pm 0.04\;.
\label{nonunitarityupperlimit}
\end{equation}
 The above result, even though consistent with zero, provides an upper experimental limit on the VLQ nonunitarity model parameter $U^{sb}$.

\section{Wilson coefficients}
The effective Hamiltonian for rare FCNC $b \to s$ transitions in the SM is derived using the operator product expansion and renormalization group evolution to integrate out heavy degrees of freedom (e.g., \(W\), \(Z\), top quark). It underlies processes like \(b \to s \gamma\), \(b \to s \ell^+ \ell^-\) (with charged leptons \(\ell = e, \mu, \tau\)), and \(b \to s \nu \bar{\nu}\).

The general form including tree-level, penguin and box contributions is:

\begin{equation}
\mathcal{H}_{\rm eff} = -\frac{4G_F}{\sqrt{2}}  V_{tb}V_{ts}^* \left\{ \sum_{i=1}^{10} C_i(\mu) O_i+C_L(\mu )O_L\right\} + \textrm{H.C.}
\label{effectiveHamiltonian}
\end{equation}

Here, \(G_F\) is the Fermi constant, \(C_i(\mu)\) and \(C_L(\mu)\) are Wilson coefficients evaluated at scale \(\mu \sim m_b\) (bottom quark mass), and \(O_i\), $O_L$ are local dimension-6 operators. Indeed, $O_7$ is responsible for $b\to s\gamma$, $O_7$, $O_9$ and $O_{10}$ for $b\to s\ell^+\ell^-$ and $O_L$ for $b\to s \nu\bar\nu$ decays.   They are given as:
\begin{align}
  O_7 &= \frac{e m_b}{16\pi^2} (\bar s \sigma^{\mu\nu} P_R b) F_{\mu\nu} \nonumber\\
  O_9 &= \frac{e^2}{16\pi^2} (\bar s \gamma^\mu P_L b) (\bar \ell \gamma_\mu \ell) \nonumber\\
  O_{10} &= \frac{e^2}{16\pi^2} (\bar s \gamma^\mu P_L b) (\bar \ell \gamma_\mu \gamma^5 \ell)\nonumber\\
  O_L &= \frac{e^2}{16\pi^2} (\bar s \gamma^\mu P_L b) (\bar \nu \gamma_\mu (1 - \gamma^5) \nu)\;,
  \label{operators}
\end{align}
 where \(P_{L,R} = (1 \mp \gamma^5)/2\), \(e\) is the electromagnetic coupling, and \(F_{\mu\nu}\) is the photon field strength.

\begin{figure}
    \centering
    \includegraphics[width=0.7\linewidth]{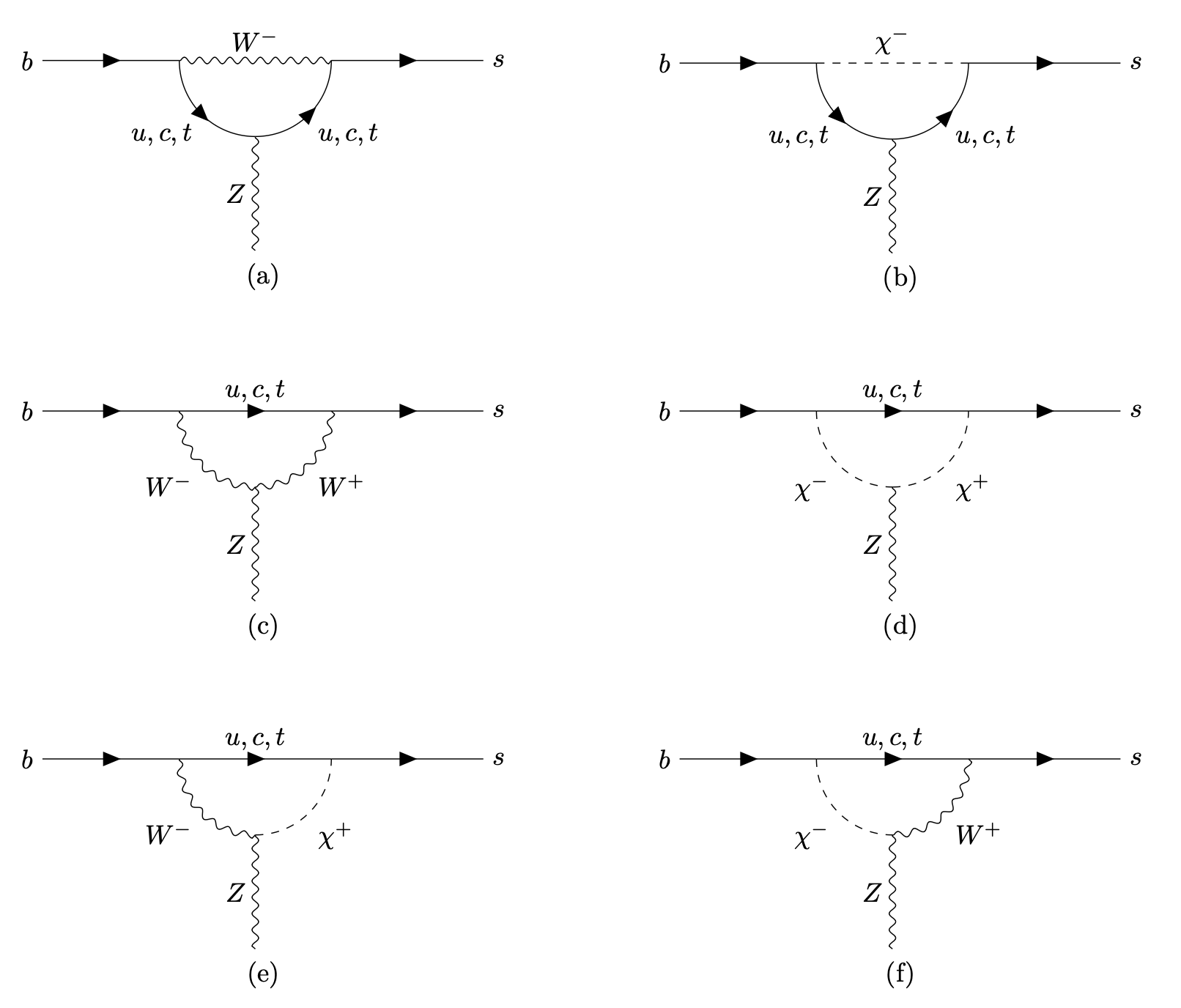}
    \caption{Penguin diagrams contributing to $b\to s Z$ FCNC process in SM.}
    \label{fig:bszpenguin}
\end{figure}

\begin{figure}
    \centering
    \includegraphics[width=0.7\linewidth]{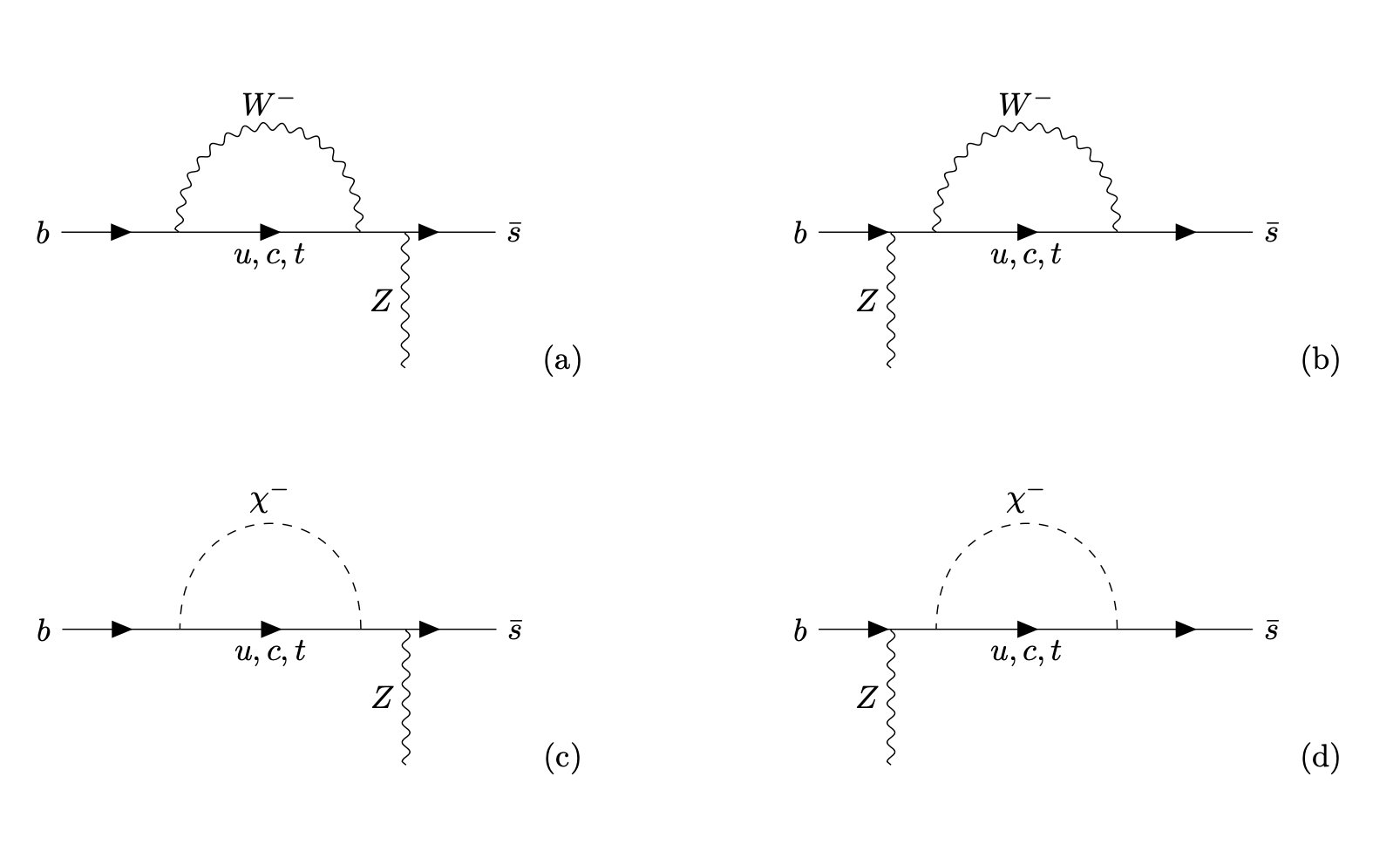}
    \caption{Self-energy diagrams contributing to $b\to s Z$ FCNC process in SM. }
    \label{fig:bszselfenergy}
\end{figure}

The Wilson coefficients \(C_i\) encode short-distance physics and are calculable in perturbation theory.  At the scale $\mu\sim M_W$, they can be written in terms of the Inami-Lim functions \cite{Inami:1980fz}:
\bea
C_7 &=& -\frac{1}{2}D_\circ^\prime (x_t) \;,\\
C_9 &=& \frac{Y_0(x_t)}{\sin^2{\theta_W}}-4Z_0(x_t)\;,\\
C_{10} &=& -\frac{Y_0(x_t)}{\sin^2{\theta_W}}\; ,\\
C_L &=& -\frac{X_0(x_t)}{\sin^2{\theta_W}}\; .
\label{wcsm}
\eea
where $x_t=m_t^2/M_W^2$ and the gauge invariant effective vertex functions are:
\bea
D_0^\prime (x_t) &=& -\frac{8x_t^3 + 5x_t^2 - 7x_t}{12(1 - x_t)^3} + \frac{x_t^2 (2 - 3x_t)}{2(1 - x_t)^4} \ln x_t , \\
X_0(x_t) &=& \frac{x_t}{8} \left( \frac{x_t + 2}{x_t-1} + \frac{3x_t - 6}{(x_t-1)^2} \ln x_t \right),  \\
Y_0(x_t) &=& \frac{x_t}{8} \left( \frac{x_t - 4}{x_t-1} + \frac{3x_t}{(x_t-1)^2} \ln x_t \right),  \\
Z_0(x_t) &=& -\frac{1}{9} \ln x_t
+ \frac{18x_t^4 - 163x_t^3 + 259x_t^2 - 108x_t}{144(x_t-1)^3} \notag \\
\quad &+& \frac{32x_t^4 - 38x_t^3 - 15x_t^2 + 18x_t}{72(x_t-1)^4} \ln x_t.
\label{effectivevertexfunctions}
\eea
We note that $X_0$ and $Y_0$ resulted from the combinations of Z-penguin and box diagrams while $D_0^\prime$ is calculated from the magnetic component of the photon penguin and $Z_0$ is a combination of the Z-penguin and electric component of the photon penguin.  It is important to point out that the calculation of the one-loop penguin and box diagrams representing $b\to s$ transition in the SM, which are dominated by the internal top-quark, include some $x_t$ independent terms which cancel out due to the unitarity of the CKM quark mixing when summation over all up-type internal quarks is carried out. Figs \ref{fig:bszpenguin} and \ref{fig:bszselfenergy} show the penguin and self-energy diagrams that contribute to $b\to sZ$ effective vertex function.

The short-distance QCD corrections to weak
decays, that is the contribution of hard gluons at energies of the order $\mathcal{O}(m_W)$ down to
hadronic scales $\mathcal{O}(1~GeV)$, can be treated in the renormalization group (RG) improved
perturbation theory \cite{Buras:2020xsm}. Indeed, the relevant scale for the effective Hamiltonian (\ref{effectiveHamiltonian}) for rare B decays ($\Delta B=1$) is $\mu=\mathcal{O}(m_b)$ and Wilson coefficients $C_i(\mu_b=\mathcal{O}(m_b))$ can be obtained from $C_i(\mu_W=\mathcal{O}(M_W))$ via the evolution matrix.
\begin{equation}
    C_i(\mu_b)=U_{ij}(\mu_b,\mu_W)C_j(\mu_W)\;.
    \label{evolution}
\end{equation}

In practice, we take $U_{ij}(\mu_b ,M_W)\to\delta_{ij} \eta_i$ (no summation over $i$) where the numerical values of $\eta_i$ are given in Table[\ref{tab:table}].

\section{VLQ contributions to Wilson Coefficients}
The addition of a single down-type VLQ leads to three distinct contributions to the Wilson coefficients. First, the nonunitarity of the $3\times 4$ quark mixing matrix leads to non-zero FCNC transition at the tree level.  Fig \ref{fig:tree} shows the tree-level Feynman diagram for $b\to s Z$ transition which is proportional to $U_{sb}$ leading to the following shifts in $C_9$, $C_{10}$ and $C_L$:
\begin{figure}
    \centering
    \includegraphics[width=0.7\linewidth]{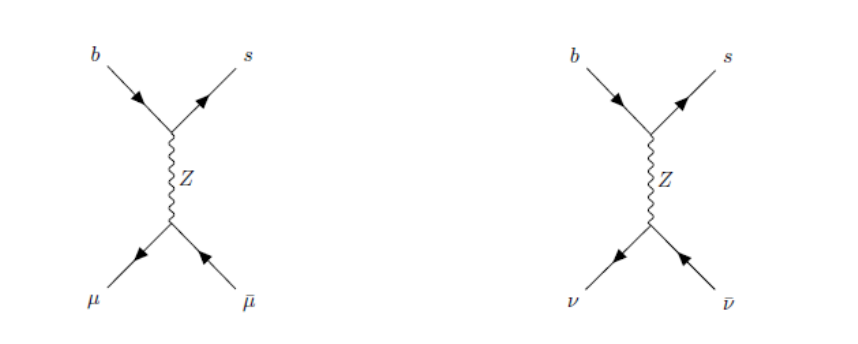}
    \caption{Tree-level contributions to Wilson coefficients $C_9$, $C_{10}$ and $C_L$.}
    \label{fig:tree}
\end{figure}
\bea
C^{\rm tree}_9 &=& \frac{\pi}{\alpha}\frac{U^{sb}}{V_{tb}V_{ts}^*}(-1+4\sin^2{\theta_W})\;,\\
C^{\rm tree}_{10} &=& \frac{\pi}{\alpha}\frac{U^{sb}}{V_{tb}V_{ts}^*}\; ,\\
C^{\rm tree}_L &=& \frac{\pi}{\alpha}\frac{U^{sb}}{V_{tb}V_{ts}^*}\; .
\label{wctree}
\eea
Second, the mass-independent terms have to be considered in the effective vertex functions (\ref{effectivevertexfunctions}) as they do not cancel out due to non-unitarity of the quark mixing matrix.  This leads to additional contributions to the Wilson coefficients:
\bea
C^{\rm NU}_{7} &=& \frac{U^{sb}}{V_{tb}V_{ts}^*}\frac{23}{36}\; ,\\
C^{\rm NU}_9 &=& -\frac{U^{sb}}{V_{tb}V_{ts}^*}\frac{1}{4\sin^2{\theta_W}}\;,\\
C^{\rm NU}_{10} &=& \frac{U^{sb}}{V_{tb}V_{ts}^*}\frac{1}{4\sin^2{\theta_W}}\; ,\\
C^{\rm NU}_L &=& \frac{U^{sb}}{V_{tb}V_{ts}^*}\frac{1}{4\sin^2{\theta_W}}\; .
\label{wcnonunitarity}
\eea

The third contribution is due to one-loop Higgs (and unphysical neutral Higgs) exchange (Fig \ref{fig:Higgs}).  Due to the fact that the Higgs coupling to fermions is proportional to their masses, these loop contributions are not suppressed by the large $M_D$.  Explicitly:  

\begin{figure}
    \centering
    \includegraphics[width=0.7\linewidth]{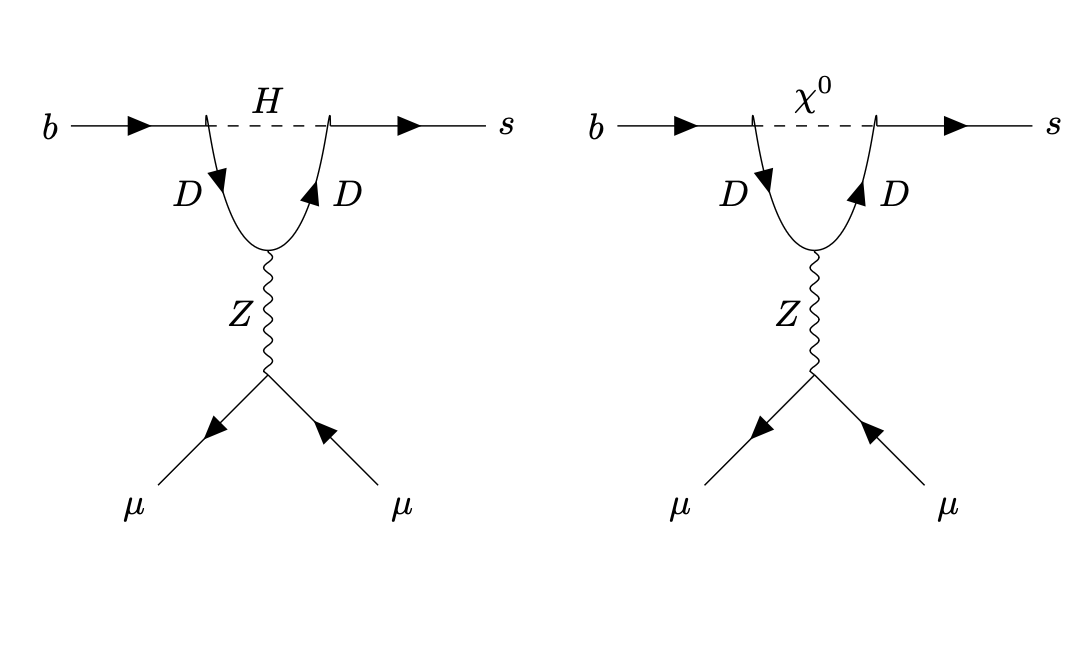}
    \caption{One-loop (unphysical) Higgs contribution to FCNC process $b\to sZ$.}
    \label{fig:Higgs}
\end{figure}

\bea
C^{\rm Higgs}_{7} &=& \frac{U_{sb}}{V_{tb} V_{ts}^*}\frac{Q_d}{8}\Bigg[ x_2\left(\frac{(2-x_2)\log{x_2}}{(1-x_2)^4}-\frac{5x_2^3}{6(1-x_2)^4}+\frac{4x_2^2}{(1-x_2)^4}-\frac{13x_2}{2(1-x_2)^4}+\frac{10}{3(1-x_2)^4}\right)\nonumber \\
&+&x_1\left(\frac{(-2+3x_1)\log{x_1}}{(1-x_1)^4}+\frac{7x_1^3}{6(1-x_1)^4}-\frac{6x_1^2}{(1-x_1)^4}+\frac{15x_1}{2(1-x_1)^4}-\frac{8}{3(1-x_1)^4}\right)\Bigg]\; ,\\
C^{\rm Higgs}_9 &=& -\frac{U^{sb}}{V_{tb} V_{ts}^*}\frac{M_D^2}{M_W^2}\left[\frac{1}{48}(F(x_1)+F(x_2))
    +\left(\frac{\frac{-1}{2}+\frac{1}{3}\sin^2{\theta_W}}{16\sin^2{\theta_W}}\right)\left(F^{'}(x_1)+F^{'}(x_2)\right)\right](4\sin^2{\theta_W}-1)\;, \nonumber \\
  {\theequation} \\
C^{\rm Higgs}_{10} &=& -\frac{U^{sb}}{V_{tb} V_{ts}^*}\frac{M_D^2}{M_W^2}\Bigg[\frac{1}{48}(F(x_1)+F(x_2))+\left(\frac{\frac{-1}{2}+\frac{1}{3}\sin^2{\theta_W}}{16\sin^2{\theta_W}}\right)\left(F^{'}(x_1)+F^{'}(x_2)\right)\Bigg]\; ,\\
C^{\rm Higgs}_L &=& -\frac{U^{sb}}{ V_{tb}V_{ts}^*} \frac{M_D^2}{M_W^2}  \Bigg[ \frac{1}{48} (F(x_1)+F(x_2))+\left(\frac{\frac{-1}{2}+\frac{1}{3}\sin^2{\theta_W}}{16 \sin^2{\theta_W}}\right) \left(F^{'}(x_1)+F^{'}(x_2)\right)\Bigg]\; ,
\label{wchiggsloop}
\eea
where
\begin{equation}
    F(x)=\left(-{\log{\frac{\mu^2}{M_D^2}}}-\frac{1}{2}+\frac{x^{2} \log x}{(1-x)^{2}}+\frac{x}{(1-x)}\right)\;,
\end{equation}

\begin{equation}
    F^{'}(x)= \left(\frac{1}{2}{\log{\frac{\mu^2}{M_D^2}}}
    +\frac{3}{4}-\frac{x^{2} \log x}{2(1-x)^{2}}-\frac{1}{2(1-x)}\right)\;,
\end{equation}
and $x_1=\frac{M_D^2}{M_H^2}$, $x_2=\frac{M_D^2}{M_Z^2}$ and $\mu$ is the renormalization scale.  In what follows, we account for the devolution of this scale from $M_W$ to $m_b$ using a multiplicative factor $\eta$.

In what follows, we shall study the interference pattern resulting from the superposition of the SM amplitude and the new VLQ amplitudes while scanning the parameter space of the VLQ. As we shall see, the measured branching ratio of $B_s\to\mu^+\mu^-$, $B\to X_s\mu^+\mu^-$ and $B\to X_s\gamma$, which are in agreement with the SM, allow us to find a window for $M_D$ which is accessible via direct searches at the LHC. We shall also investigate these effects of the VLQ on the recently measured $B\to K\nu\bar\nu$ anomaly.

\section{FCNC rare B$_{(s)}$ decays}

The branching ratio of $B_s\to\mu^+\mu^-$ which provides the most stringent constraints on the VLQ parameters is given as:
\begin{equation}
      \mathcal{BR}(B_s \to \mu^+ \mu^-) 
= \tau_{B_s} \, \frac{G_F^2}{\pi} 
\left( \frac{\alpha}{4\pi } \right)^2 
f_{B_s}^2 \, M_{B_s} \, m_\mu^2 
\sqrt{1 - \frac{4 m_\mu^2}{M_{B_s}^2}}\
\left| V_{tb}^* V_{ts} \right|^2 
\left|\eta_{10} C_{10} \right|^2 \;,
\label{Bsmumu}
\end{equation}
where $\tau_{B_s}$, $f_{B_s}$ and $m_{B_s}$ are the life-time, decay constant and the mass of $B_s$-meson and $m_\mu$ is the muon mass.  The numerical values of all parameters are given Table [\ref{tab:table}].

For the inclusive decays, it is customary to normalize the branching ratios by using inclusive semileptonic $B\to X_ce\bar \nu$ in order to reduce the sensitivity to the b-quark mass ($m_b$) input.  Accordingly, the branching ratio for the inclusive $B\to X_s\mu^+\mu^-$ can be expressed as:
\begin{equation}
    \mathcal{BR}(B \to X_s \mu^+ \mu^-) = \frac{\alpha^2\, \mathcal{BR}(B \to X_c e \bar{\nu})}{4\pi^2\, f(\hat{m}_c)\, \kappa(\hat{m}_c)} 
    \cdot \frac{|V_{ts}^* V_{tb}|^2}{|V_{cb}|^2} 
    \int D(z)\, dz,
    \label{BR}
\end{equation}
where 
\begin{equation}
  D(z) = (1 - z)^2 \Bigg[(1 + 2z) \left( |C_9(\mu_b)|^2 + |C_{10}(\mu_b)|^2 \right)
+ 4 \left(1 + \frac{2}{z} \right) |C_7(\mu_b)|^2
+ 12\, \text{Re}\left(C_7(\mu_b) {C_9}^*(\mu_b) \right) \Bigg] \;,
\end{equation}
with $z \equiv \frac{q^2}{m^2_ b} \equiv \frac{(p_{µ^+} + p_{µ^-} )^2}{m^2_b}$.  $f(\hat{m_c})$ is the phase-space factor in $BR((B\rightarrow X_c e \bar{\nu})$ and $\kappa(\hat{m_c})$ is the QCD correction of one-loop given in \cite{Kim:1989ac},
 \begin{equation}
     f(\hat{m_c})=1-8\hat{m_c}^2+8\hat{m_c}^6-\hat{m_c}^8-24\hat{m_c}^4\ln{\hat{m_c}}
     \label{fmc}
\end{equation}
\begin{equation}
    \kappa(\hat{m_c})=1-\frac{2\alpha_s(m_b)}{3\pi}\left((\pi^2-\frac{31}{4})(1-\hat{m_c})^2+\frac{3}{2}\right)
    \label{kmc}
 \end{equation}  
  where $\hat{m}_c = \frac{m_c}{m_b}$.  To compare with data, the branching ratio (\ref{BR}) is integrated in the low-$q^2$($1~\text{GeV}^2 \leq q^2 \leq 6~\text{GeV}^2$) and the high-$q^2$($14.4 ~\text{GeV}^2 \leq  q^2 \leq  m^2_b$ ) regions where the experimental results are given as \cite{Belle:2005fli}:
\begin{align}
    \mathcal{BR}(\bar{B} \to X_s \mu^+ \mu^-)_{\text{low}} =(1.60 \pm 0.50) \times 10^{-6}\;,\\ \quad  \mathcal{BR}(\bar{B} \to X_s \mu^+ \mu^-)_{\text{high}} =(0.44 \pm 0.12) \times 10^{-6}\;.
\end{align}
Similarly, the branching ratio for inclusive radiative decay $B\to X_s\gamma$ is written as
\be
\mathcal{BR}(B\to X_s\gamma)
= \mathcal{BR}(B \to X_c e \bar{\nu})\;
\frac{6\,\alpha}{\pi}\;
\frac{|V_{tb}V_{ts}^*|^2}{|V_{cb}|^2}\;
\frac{|C_7(\mu_b)|^2}{f(\hat{m_c}^2)}\;.
\label{radiativedecay}
\ee

As for the exclusive $B \to K \nu \bar{\nu}$ decay, the differential branching ratio can be written as:

\begin{align*}
     \frac{dB(\mathcal{BR}\rightarrow K \nu \bar{\nu})}{dq^2}=\frac{G_F^2\alpha^2(M_Z)}{32\pi^5}{|C_L(\mu_b )|}^2\tau_B |V_{ts}^* V_{tb}|^2 |\vec {p}_K|^3 {f}^2_+(q^2)
     \label{dBR-Knunu-NP}
     \end{align*}  
where $\vec{p}_K$ is the 3-momentum of the K-meson and $f_+$ is one of the form factors parametrizing the hadronic $B\to K$ transition:
\begin{equation}
\langle K(p_K)|\bar{s}\gamma^\mu b|B(p_B)\rangle
= f_+(q^2)
\left[
(p_B^\mu + p_K^\mu )
-
\frac{m_B^2 - m_K^2}{q^2}\,q^\mu
\right]
+ f_0(q^2)
\frac{m_B^2 - m_K^2}{q^2}\,q^\mu\,,
\label{formfactor}
\end{equation}
and \(q^\mu = p_B^\mu - p_K^\mu\) is the 4-momentum transfer.  We use a widely utilized parametrization of $f_+$:
\begin{align}
    f_+(q^2) = \frac{r_1}{1 - \left(\frac{q}{M}\right)^2} + \frac{r_2}{\left(1 - \left(\frac{q}{M}\right)^2\right)^2},
\end{align}

with $r_1=0.162$, $r_2=0.173$ and $M=5.41$ GeV \cite{Ball:2004ye}

\begin{table}[h!]
	\centering
	\resizebox{\textwidth}{!}{%
		\begin{tabular}{|c|c|c|c|}
			\hline
			Parameter & Value  & Parameters & Value \\ \hline
			$\alpha^{-1}( M_W)$& $130.3\pm2.3$ &  $M_Z$ & $91.1876\pm0.0021$ GeV \\ \hline
			$\alpha^{-1}(m_b)$ & $133.06 \pm 0.01 $ & $M_W$ & $80.3779 \pm 0.024$ GeV \\ \hline
			$\alpha_s (M_Z)$& $0.118100$  & $G_F$ & $1.16 \times 10^{-5}\, \text{GeV}^{-2}$ \\ \hline
			$V_{cb}$ & $.04197\pm 0.0048$ & $M_{B_s}$ & $5.367 \pm 0.0024$ GeV/$c^2$ \\ \hline
			$ f_{B_s}$ & $ 0.230$&$M_{\mu}$ & $105.658$ MeV/$c^2$\\ \hline
			$\frac{m_c}{m_b}$& $0.29\pm 0.02$& $M_{K}$ & $0.493677$ GeV/$c^2$\\ \hline
			$ \tau_{B_s}$& $1.515\pm0.006$& $m_b(m_b)$ & $~ 4.186 \pm 0.006$ GeV \\ \hline
			$\eta_{7}$ & $0.64$ &$M_{H}$ & $125$ GeV/$c^2$\\ \hline
			
			$\eta_{9}$& $0.89$& $M_{B}$ & $5.27915$ GeV/ $c^2$\\ \hline
			$\eta_{10}$&$1.0113$& $ \tau_{B}$& $1.638\pm0.004$\\ \hline
			$\eta_{L}$ & $0.995$& $ m_t(m_t)$ & $ 172.57 \pm 0.0.29$ GeV\\ \hline
			$X_0$ & $1.48^{+0.07}_{-0.05}$ & $V_{ts}^* V_{tb}$& $0.0403 \pm 0.0009$\\ \hline
			$X(x_t)$ & $4.09^{+0.11}_{-0.08}$ & $\mathcal{BR}(B\rightarrow X_{s}e\bar{\nu})$ & $(10.61 \pm 0.17)\times 10^{-2}$ \\ \hline
		\end{tabular}%
	}
	\caption{Numerical values of input parameters}
	\label{tab:table}
\end{table}
\section{Results}
\begin{figure}
    \centering
    \includegraphics[width=0.8\linewidth]{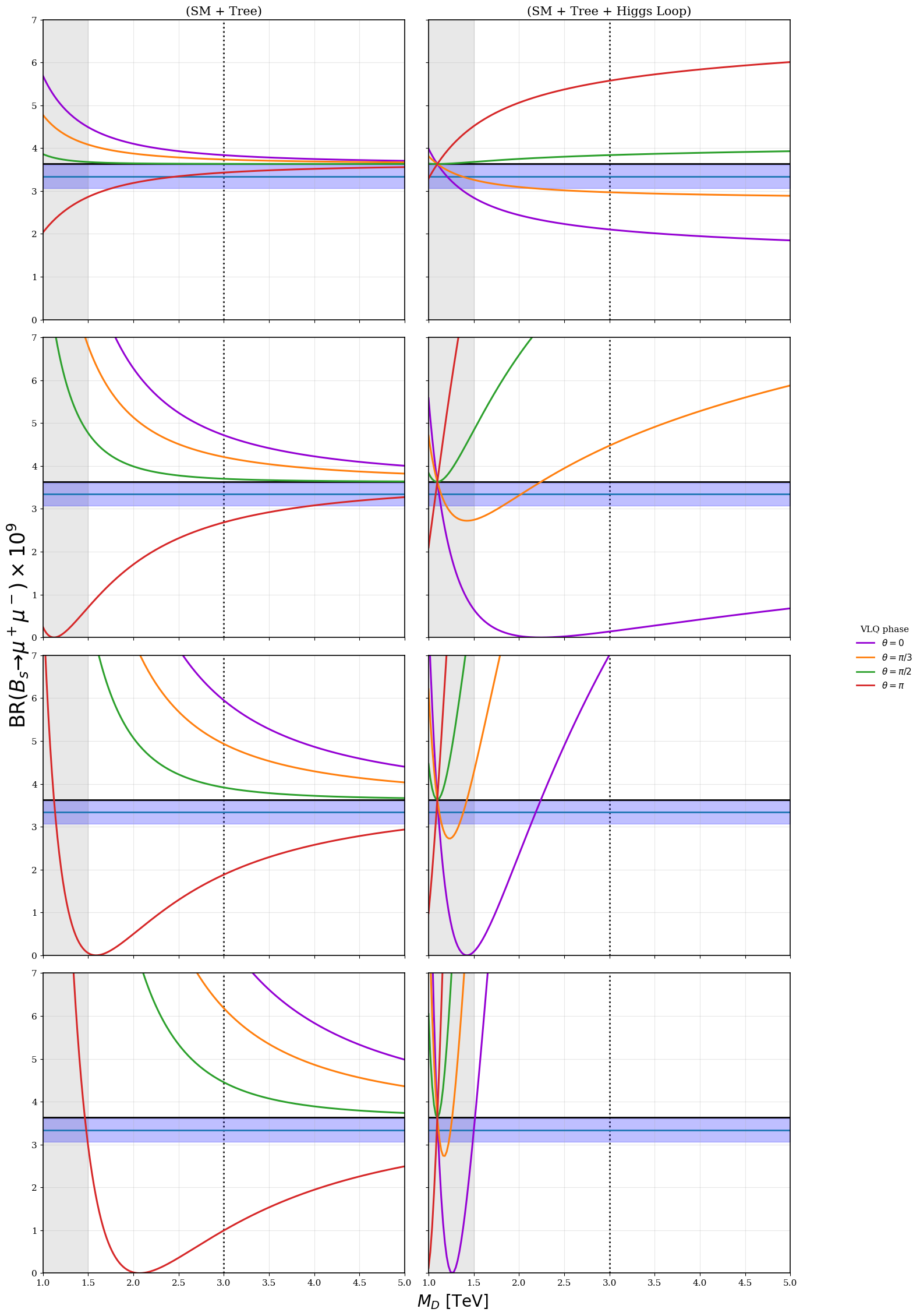}
    \caption{The branching ratio for the purely leptonic decay, $B_s \to \mu^+ \mu^-$, versus the VLQ mass, $M_D$. As can be seen, adding the Higgs loop contribution drives the VLQ mass to the accessible range for LHC direct search. }
    \label{fig:Bsmumu}
\end{figure}

\begin{figure}
    \centering
    \includegraphics[width=1.0\linewidth]{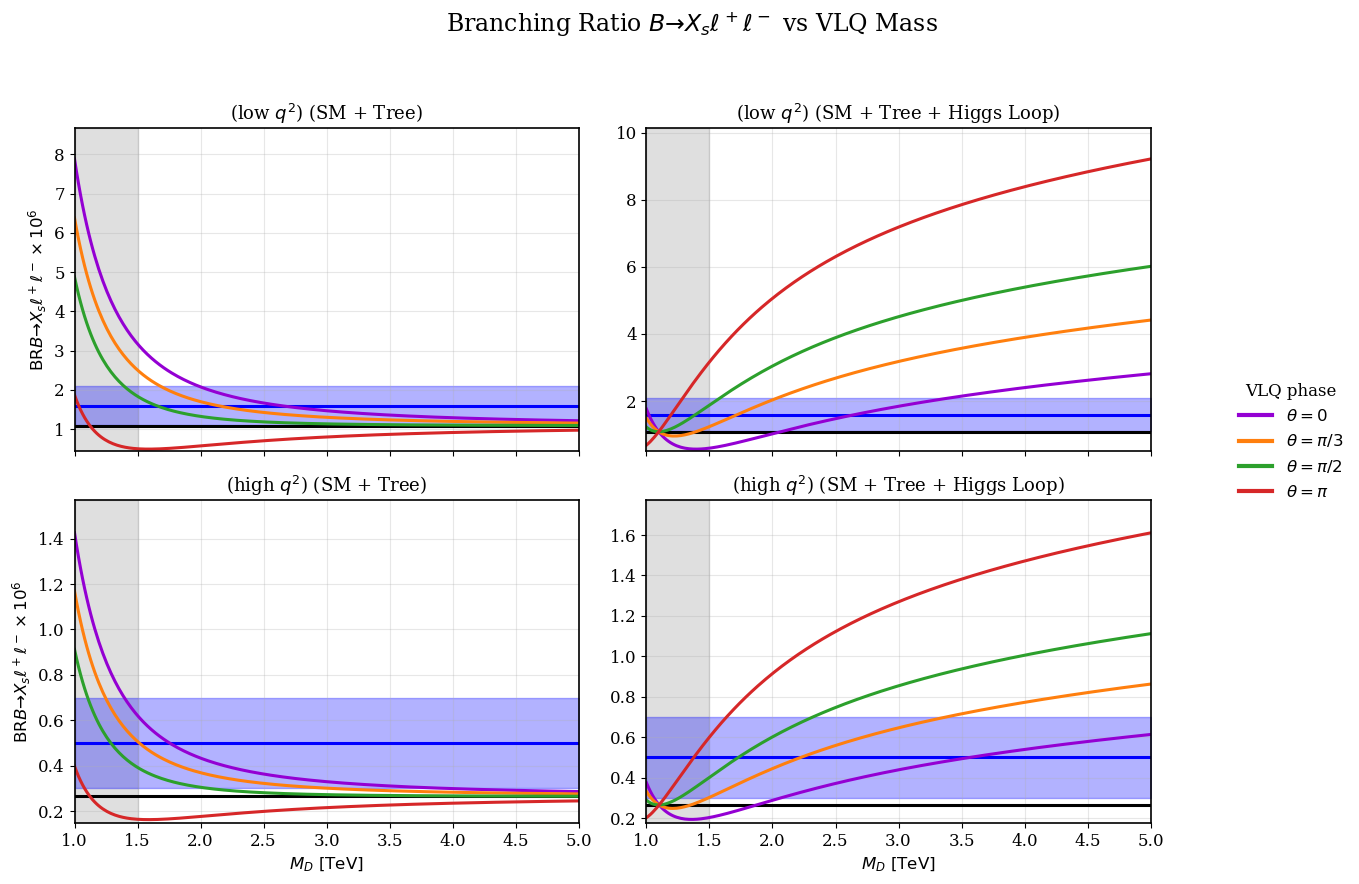}
    \caption{The branching ratio for the inclusive decay, $B \to X_s \mu^+ \mu^-$, versus the VLQ mass, $M_D$. }
    \label{fig:BXsmumu}
\end{figure}

\begin{figure}
    \centering
    \includegraphics[width=1.0\linewidth]{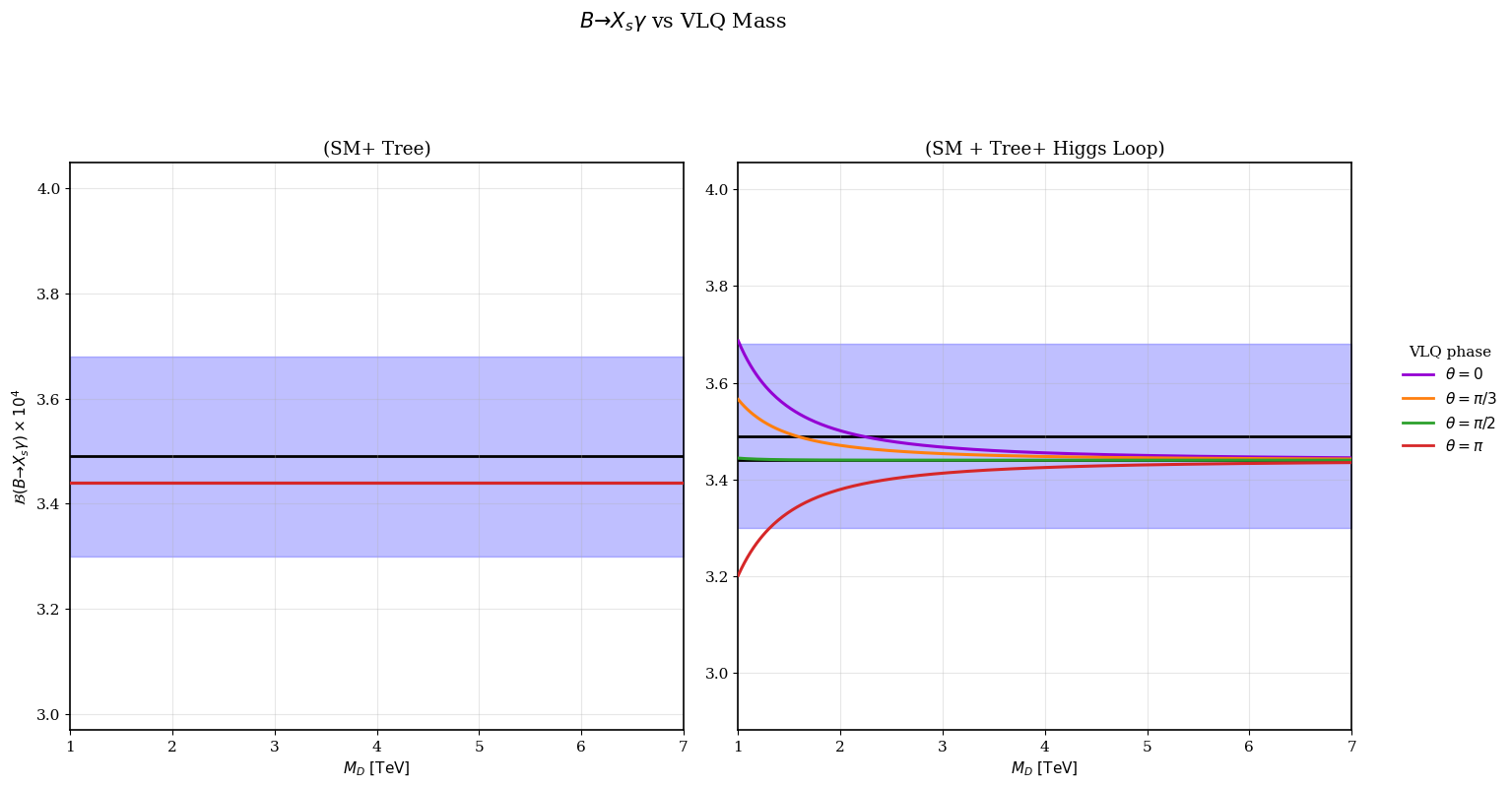}
    \caption{Branching ratio for $B \to X_s \gamma$. There is no tree-level VLQ amplitude for this decay.}
    \label{fig:BXsgamma}
\end{figure}

Our predictions for the branching ratios for $B_s\to \mu^+\mu^-$, $B\to X_s \mu^+\mu^-$ and $B\to X_s\gamma$ vs $M_D$ (with $y_{sb}$ fixed) are shown in Figs. \ref{fig:Bsmumu}, \ref{fig:BXsmumu} and \ref{fig:BXsgamma}. In all figures, the vertical grey band indicates the values of $M_D$ already excluded by direct searches at the LHC (for definiteness, we take the exclusion region to be $M_D \le 1.5$ TeV) and  the horizontal blue band is the PDG average of measurements for the branching ratio.  The left columns represent the effect of the tree-level VLQ contribution whereas the right columns show the effects of all three VLQ contributions. For each row the magnitude of $y_{sb}$ is fixed: $y_{sb}=0.00165$ for the first row, $y_{sb}=0.00826$ for the second row, $y_{sb}=0.0165$ for the third row and $y_{sb}=0.02809$ for the last row.  The curves in each plot represent different values of $\theta_{sb}$: red ($\theta_{sb}=\pi$), green ($\theta_{sb}=\pi/2$), orange ($\theta_{sb}=\pi/3$) and purple ($\theta_{sb}=0$). For completeness, we also account for the breakdown of the GIM mechanism due to CKM non-unitarity, but this correction is numerically small and does not affect our conclusions.

As pointed out in \cite{Morozumi:2018cnc}, the decay $B_s \to \mu^+ \mu^-$ provides the most stringent constraint on the singlet down type VLQ scenario. Our results for this decay are shown in Figure~\ref{fig:Bsmumu}.  The left hand column shows that, for any value of $\theta_{sb}$, the predicted branching ratios approach asymptotically the SM prediction, as expected with only the tree-level VLQ contribution. Note that, for $y_{sb} \simeq 1$, we find $M_D > 5.0~\text{TeV}$, consistent with the finding in Ref.~\cite{Morozumi:2018cnc}.  The right-hand column shows the impact of including the Higgs--VLQ loop. The first notable change is that all the curves now intersect with the SM prediction, independent of the phase $\theta$, for $M_D \simeq 1.2$ TeV. This is the value of $M_D$ for which the tree and Higgs--VLQ loop amplitudes cancel each other exactly. However, $M_D \simeq 1.2$ TeV cannot explain the agreement between data and the SM with a VLQ since it lies in the current exclusion region via direct searches. The second notable feature is that each curve re-intersect the SM prediction at a higher value of $M_D$, due to interference between all contributing amplitudes, the maximum of which, $M_D^\text{max}$, being determined by the $\theta=0$ curve. As can be seen, $M_D^\text{max}$ moves towards the LHC's discovery reach as $y_{sb}$ increases. The third row is for the representative value, $y_{sb}=0.0165$,  with $M_D^{\rm max} \simeq 2.2~\text{TeV}$ which is clearly within the LHC's discovery reach. For $y_{sb}=0.02809$, $M_D^\text{max}$ has already entered the region excluded by current direct searches. This means that higher values of $y_{sb}$ are excluded even though they do not yet saturate the CKM unitarity bound given by Eq. \eqref{nonunitarity}. 

Having identified the representative values of $y_{sb}=0.0165$ and $M^{\rm max}_D=2.2$ TeV, it is crucial to verify that they do not affect the  branching ratios for the inclusive decays, $B \to X_s \mu^+ \mu^-$ and $B \to X_s \gamma$, for which the SM  agrees with experiment. In figure \ref{fig:BXsmumu}, we show that this is the case for both the low and high $q^2$ bins of $B \to X_s \mu^+ \mu^-$. For $B \to X_s \gamma$, there is no tree-level amplitude and consequently, this process does not yield an upper bound on $M_D$. Nevertheless, for $y_{sb}=0.0165$ and $M_D \simeq 2.2$ TeV, our prediction for branching ratio agrees with experiment as shown in Fig. \ref{fig:BXsgamma}. 

\begin{figure}
    \centering
    \includegraphics[width=1.0\linewidth]{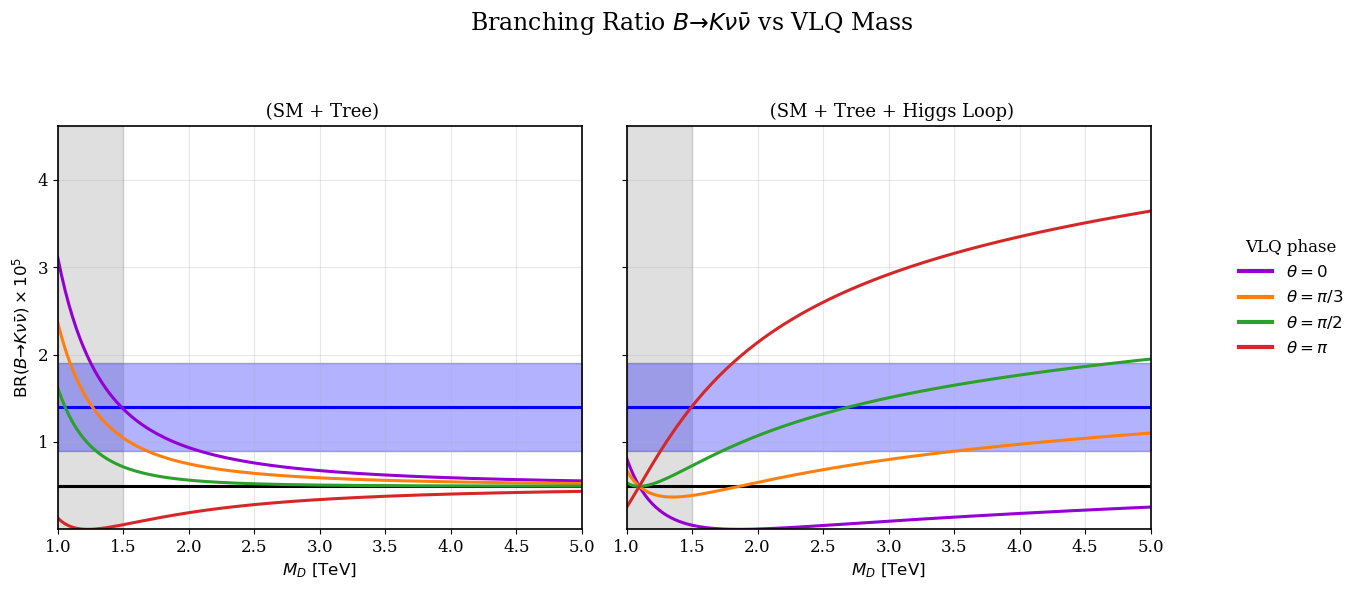}
    \caption{The branching ratio for $B \to K \nu \bar{\nu}$ versus the VLQ mass $M_D$. The vertical grey band is excluded by direct searches at the LHC and the horizontal blue band is the  BELLE II measured branching ratio. Note the difference in phase dependence compared to the previous decay modes.}
    \label{fig:Bknunu}
\end{figure}

Finally, we explore the implications of this single VLQ scenario for the rare $B \to K \nu \bar{\nu}$ decay for which there is currently a tension between the SM and the Belle II measurement \cite{Belle-II:2023esi}. Our results are shown in Fig. \ref{fig:Bknunu}. As for the other decays, the curves intersect with the SM prediction at $M_D \simeq 1.2$ TeV which is already excluded by direct searches. On the other hand, the purple curve ($\theta=0$) does not re-enter the experimental band for $M_D=2.2$ TeV. Instead, it is the red ($\theta=\pi$) curve that does so. This means that we are unable to describe simultaneously the $B_s \to \mu^+ \mu^-$, $B \to X_s \mu^+ \mu^-$ and $B \to X_s \gamma$ as well as the $B \to K \nu \bar{\nu}$ branching ratios. Therefore, our minimal VLQ scenario does not alleviate the current tension between the SM and experiment for $B \to K \nu \bar{\nu}$, unless one is prepared to assume a different phase only for this process: $\theta_{sb}=\pi$ for $B \to K \nu \bar{\nu}$ while $\theta_{sb}=0$ for all the other decays. 

\section{Conclusions}
We have shown that for a minimal New Physics scenario with a singlet down type vector-like quark, interference effects involving the previously overlooked Higgs-VLQ loop amplitude implies an upper bound on $M_D$ while remaining in agreement with both the SM and the data for $B_s\to \mu^+\mu^-$ and $B\to X_s (\gamma, \mu^+\mu^- )$. Furthermore, for a safely allowed violation of the CKM unitarity, this upper bound lies within the discovery reach at the LHC. This means that if the singlet down type VLQ is not found at the LHC, this minimal NP scenario can be ruled out.  The minimal scenario considered here cannot simultaneously describe the data for $B_s \to \mu^+ \mu^-$ and $B \to X_s (\gamma,\mu^+ \mu^-)$ while explaining the current anomaly in $B \to K \nu \bar{\nu}$.  
\bibliographystyle{apsrev}
\bibliography{ref}

\end{document}